\documentclass[aps,prl,twocolumn,nofootinbib,tightenlines,showpacs]{revtex4}

\usepackage{epsfig,bm}

\begin{document}

\preprint{hep-ph/0111132}

%%%%%%%%%%%%%%%%%%%%% Title %%%%%%%%%%%%%%%%%%%%%%

\title{
Quarkonium-Hadron Interactions in QCD}

%%%%%%%%%%%%%%%%%%%% Authors %%%%%%%%%%%%%%%%%%%%%

\author{Yongseok Oh}%
\email{yoh@phya.yonsei.ac.kr}

\author{Sungsik Kim}%
\email{sskim@phya.yonsei.ac.kr}
\altaffiliation[Present address: ]{Department of Physics,
Sungkyunkwan University, Suwon 440-746, Korea}

\author{Su Houng Lee}%
\email{suhoung@phya.yonsei.ac.kr}

\affiliation{Institute of Physics and Applied Physics, Department of
Physics, Yonsei University, Seoul 120-749, Korea}

%%%%%%%%%%%%%%%%%%%% Abstract %%%%%%%%%%%%%%%%%%%%%

\begin{abstract}
%The leading order non-relativistic QCD result for the quarkonium-hadron
%interaction cross section has been previously derived by Peskin and
%Bhanot using the operator product expansion.
%Here, we derive the same results using the QCD factorization formula
%combined with the Bethe-Salpeter amplitude for the heavy quark bound
%state.  
The leading order non-relativistic QCD result for the quarkonium-hadron
interaction cross section is derived by using the QCD factorization formula
combined with the Bethe-Salpeter amplitude for the heavy quark bound
state.  
Our result shows the equivalence of this approach with the 
operator product expansion even for the bound state scattering
to leading order in QCD.
We also apply the formula to the charmonium system and investigate the
relativistic correction coming from the phase space of the reaction.
We find that although $J/\psi$-gluon cross-section gets a large
relativistic correction at higher energies, the correction becomes
small in the $J/\psi$-hadron cross section due to the increase of the 
gluon distribution at smaller $x$ inside the hadron. 
\end{abstract}

\pacs{25.75.-q, 12.38.-t, 13.75.-n, 13.85.Lg}

\maketitle

%%%%%%%%%%%%%%%%%%%% Text %%%%%%%%%%%%%%%%%%%%%

More than two decades ago, Peskin \cite{Peskin79} and Bhanot and Peskin 
\cite{BP79} showed that one could apply  
perturbative Quantum Chromodynamics (pQCD) to calculate the 
interactions between the bound state of heavy quarks and the light hadrons. 
Such calculation was feasible because in the large quark mass limit one 
could consistently obtain the leading order operator product expansion 
(OPE) of the correlation function between two heavy meson currents in 
the light hadron state. 
The justification of such calculation stems from the fact that there exist
two relevant scales in the bound state \cite{ADM78} in the large quark
mass limit, namely the binding energy, which becomes Coulomb-like and
scales as $mg^4$, and the typical momentum scale of the bound state,
which scales like $mg^2$, where $m$ is the heavy quark mass and $g$ the
quark-gluon coupling constant.
Hence, taking the separation scale of the OPE to be the binding
energy, it is consistent to take into account the bound state
property, which is generated by the typical momentum scale of the
bound state, into the process-dependent Wilson coefficient.

The calculation is challenging in itself, and has attracted recent
interests because when applied to the $J/\psi$ system \cite{KS94,AGGA01},
the strength of the $J/\psi$-hadron inelastic scattering cross section
determines the amount of the $J/\psi$ suppression
\cite{MS86} in heavy ion collisions due to hadronic final state
interactions \cite{Vogt99}.
Moreover, other existing model calculations for $\sigma_{J/\psi-h}$,
based on meson exchange models \cite{meson-ex}, quark exchange models
\cite{WSB01}, or QCD sum rules \cite{NNMK02} give very different energy
dependence and magnitude near threshold, which shows the importance of
the nonperturbative effects.
Therefore, more careful analyses of each calculation are necessary,
where the limitations and expected corrections would be explicitly
spelled out.

In this paper, we will derive the leading order pQCD result using the QCD
factorization theorem.
The result obtained by Peskin and Bhanot was derived within
the OPE \cite{Peskin79,BP79}.
The equivalence between the OPE and the factorization approach is well
established in the deep inelastic scattering and Drell-Yan processes
\cite{HVN81}.
As will be shown here, the equivalence between the two approaches are
also true for bound state scattering to leading order in QCD.
Similar approach has been used by one of us (SHL) to estimate the
dissociation cross section of the $J/\psi$ at finite temperature
\cite{HLZ88}.
The factorization formula also provides a manageable starting point to
calculate higher twist gluonic effects \cite{BB00,KL01}, which should be
nontrivial for the $J/\psi$-hadron scattering.

\begin{figure}[b]
\epsfig{file=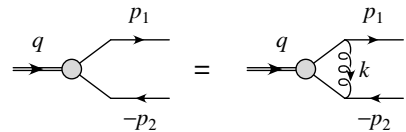, width=0.6\hsize}
\caption{The Bethe-Salpeter equation for the quarkonium $\Phi$.}
\label{fig:BS}
\end{figure}

We refer to a bound state of heavy quark and its own antiquark as $\Phi$.
According to the factorization formula, the total scattering cross section
of $\Phi$ with a hadron $h$ can be written as \cite{BP79,EFP82}
\begin{equation}
\sigma_{\Phi h}(\nu) = \int_0^1 dx \sigma_{\Phi g}(x \nu) g(x),
\label{eq:fact}
\end{equation}
with $\nu = p \cdot q / M_\Phi$, where $p$ ($q$) is the momentum of the
hadron ($\Phi$) and $M_\Phi$ is the $\Phi$ mass.
Here, $\sigma_{\Phi g}$ is the perturbative $\Phi$-gluon scattering
cross section and $g(x)$ is the leading twist gluon distribution function
within the hadron.
The separation scale is taken to be the binding energy of the bound state.
Hence, the bound state properties have to be taken into account in
$\sigma_{\Phi g}$.
This can be accomplished by introducing the Bethe-Salpeter (BS) amplitude
$\Gamma(p_1,-p_2)$, which  satisfies \cite{BS51}
%\begin{widetext}
\begin{eqnarray}
\Gamma_\mu(p_1,-p_2) &=& i C_{\rm color} \int \frac{d^4 k}{(2\pi)^4} g^2
V(k) \gamma^\nu \Delta(p_1+k)
\nonumber \\ && \hspace{-1cm} \mbox{} \times
\Gamma_\mu(p_1+k,-p_2+k) \Delta(-p_2+k)
\gamma_\nu,
\end{eqnarray}
%\end{widetext}
where $C_{\rm color} = (N_c^2-1)/(2N_c)$ with the number of color $N_c$.
The trivial color indices have been suppressed.   
Figure~\ref{fig:BS} shows the diagrammatic representation of the
BS equation, where $p_1$ ($-p_2$) is the four-momentum of
the heavy quark (anti-quark) with
$\Delta(p) = 1/(p\!\!\!/ - m + i\epsilon)$ and
$V(k) = - 1/(k^2 + i\epsilon)$.

We introduce $\phi_\mu$, which is defined as
\begin{equation}
\phi_\mu(q,{\bf p})
\equiv
\left(\frac{N_c}{M_\Phi} \right)^{1/2} \!
\int \frac{dp_0}{2\pi}
\Delta(p_1) \Gamma_\mu(p_1,-p_2) \Delta(-p_2),
\label{eq:Phimu}
\end{equation}
where $q=p_1+p_2$ and $p = (p_1-p_2)/2$.
We work in the $\Phi$ rest frame and introduce the binding energy
$\varepsilon$, such that $q_0 = M_\Phi = 2 m+ \varepsilon$  
($\varepsilon<0$).  
Then, in the non-relativistic limit, Eq. (\ref{eq:Phimu}) reduces to
\begin{eqnarray}
\phi_\mu(q,{\bf p}) &=&
\frac{-i}{\varepsilon - {\bf p}^2/m}
\sqrt{\frac{N_c}{M_\Phi}} \frac{1+\gamma_0}{2}
\nonumber \\ && \mbox{} \times
\Gamma_\mu \left(\frac{q}{2}+p,-\frac{q}{2}+p \right)
\frac{1-\gamma_0}{2},
\label{eq:phi}
\end{eqnarray}
where $p_0 = \varepsilon/2-{\bf p}^2/(2m)$.   
In this limit, the BS amplitude reduces to
\begin{eqnarray}
&& \Gamma_\mu(q/2+p,-q/2+p)
\nonumber \\ 
&=&
- \left( \varepsilon - \frac{{\bf p}^2}{m} \right) \sqrt{\frac{M_\Phi}{N_c}}
\psi({\bf p})
\frac{1+\gamma_0}{2} \gamma_i \delta_{\mu i} \frac{1-\gamma_0}{2}, 
\label{eq:BS}
\end{eqnarray}
and the corresponding BS equation becomes the non-relativistic
Schr\"odinger equation for the Coulombic bound state,
\begin{equation}
\left( \varepsilon - \frac{{\bf p}^2}{m} \right) \psi({\bf p}) = -g^2
C_{\rm color}
\int \frac{d^3 k}{(2\pi)^3} V({\bf k}) \psi({\bf p}+{\bf k}),
\label{BSeq}
\end{equation}
so that $\psi({\bf p})$ is the normalized wave function for the bound state.  
For the pseudo scalar bound state, we replace $\gamma_i \delta_{\mu i}$ to
$\gamma_5$ in Eq. (\ref{eq:BS}), but obtain the same spatial wave function.
(Some discussions on the realistic potential can be found, for example,
in Ref.~\cite{EFG00}.)
It is straightforward to verify that our formalism gives the well-known
result for the $\Phi \to e^+ e^-$ decay width,
$
\Gamma(\Phi \to e^+ e^-) =
\frac{16 \pi \alpha_{\rm em}^2 Q_h^2 N_c}{3M_\Phi^2}
\left| \psi(r=0) \right|^2,
$
where $\alpha_{\rm em} = e^2/(4\pi)$ and $Q_h$ is the heavy quark charge
in the unit of $e$.

\begin{figure}[b]
\epsfig{file=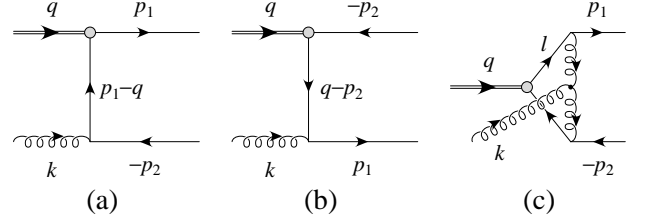, width=0.95\hsize}
\caption{Scattering processes of $\Phi$ and gluon.}
\label{fig:reac}
\end{figure}

With the BS amplitude defined as Eq.~(\ref{eq:BS}), we now obtain the
$\Phi$-gluon scattering amplitude with the processes depicted in
Fig.~\ref{fig:reac}.
The scattering matrix elements can be written as 
$\mathcal{M} = \varepsilon_\mu(\Phi) \mathcal{M}^{\mu\nu}
\varepsilon_\nu(g)$,
where $\varepsilon_\mu(\Phi)$ and $\varepsilon_\nu(g)$ are the polarization
vectors of the $\Phi$ and the gluon, respectively.
We first compute $\mathcal{M}_{1}^{\mu\nu}$, which comes from
Figs.~\ref{fig:reac}(a,b),
\begin{eqnarray}
\mathcal{M}^{\mu\nu}_{1} &=& - g \bar{u}(p_1) \bigl[ \Gamma^\mu
(p_1, -q+p_1) \Delta(-q+p_1) \gamma^\nu
\nonumber \\ && \mbox{} \!\!\!
+ \gamma^\nu \Delta(q-p_2) \Gamma^\mu(q-p_2,-p_2) \bigr] T^a v(p_2),
\label{M1}
\end{eqnarray}
where $u(p)$ and $v(p)$ are the Dirac spinors of the quark and
antiquark, respectively, $a$ is the color index of the incoming gluon,
and $T^a$ is the color matrix with
$\mbox{Tr}(T^a T^b) = \delta^{ab}/2$.  
To obtain the leading order non-relativistic result, we have to discuss 
the counting scheme in the rest frame of $\Phi$.
First the binding energy $\varepsilon_0 = m \left[ N_c g^2 / (16\pi)
\right]^2$ is $O(mg^4)$.  
Combined with the energy conservation $q+k=p_1+p_2$ in the non-relativistic
limit this implies the following counting,  
\begin{eqnarray}
|{\bf p_1}| \sim |{\bf p_2}| \sim O(mg^2),~~  k^0=|{\bf k}|\sim O(mg^4).
\label{counting}
\end{eqnarray}  
Then the leading order result of Eq.~(\ref{M1}) reads
\begin{eqnarray}
\mathcal{M}_{1}^{\mu\nu} &=&  g \sqrt{\frac{M_\Phi}{N_c}}
\biggl\{ -{\bf k} \cdot
\frac{\partial \psi({\bf p})}{\partial {\bf p}} \delta^{\nu 0}
+ \frac{2p^j}{m} \psi({\bf p}) \delta^{\nu j} \biggr\} \delta^{\mu i}
\nonumber \\ && \mbox{} \times
\bar{u}(p_1)\frac{1+\gamma_0}{2}
\gamma^i T^a \frac{1-\gamma_0}{2} v(p_2).
\label{eq:AM1}
\end{eqnarray}

Figure~\ref{fig:reac}(c) gives
\begin{widetext}
\begin{eqnarray}
\mathcal{M}_2^{\mu\nu} &=& g^3 f_{abc} T^c T^b
\bar{u}(p_1) \int \frac{d^4 l}{(2\pi)^4} V(l-p_1) V(l-p_1+k)
\gamma_\beta \Delta(l) \Gamma^\mu(l,l-q) \Delta(l-q) \gamma_\alpha v(p_2)
\nonumber \\ && \mbox{} \times
\left[ g^{\alpha\beta} (2l-2p_1+k)^\nu + g^{\beta\nu} (p_1-l+k)^\alpha -
g^{\nu\alpha} (l-p_1+2k)^\beta \right].
\label{eq:M2}
\end{eqnarray}
\end{widetext}
The color factor of this diagram is $f_{abc} T^c T^b = - \frac{i}{2}
N_c T^a$.
There is one more diagram, where the external gluon leg is attached to
the internal quark line within the internal loop.
However such a diagram carries the color factor $T^b T^a T^b = - T^a /
(2N_c)$ (color adjoint potential) and hence is suppressed by
$1/N_c^2$ compared to Fig.~\ref{fig:reac}(c) \cite{Peskin79}.

Using the BS equation and the counting (\ref{counting}), 
we get
\begin{eqnarray}
M_2^{\mu\nu} &=& - g \sqrt{\frac{M_\Phi}{N_c}}
\biggl\{ k_0 \frac{\partial \psi({\bf p})}{\partial
p^j} + \frac{2p^j}{m} \psi({\bf p}) \biggr\}
\delta^{\mu i} \delta^{\nu j}
\nonumber \\ && \mbox{} \times
\bar{u}(p_1) \frac{1+\gamma_0}{2} \gamma^i \frac{1-\gamma_0}{2} T^a v(p_2).
\label{eq:AM2}
\end{eqnarray}
Collecting Eqs.~(\ref{eq:AM1}) and (\ref{eq:AM2}) gives the (gauge invariant)
leading order result for the scattering amplitude as 
\begin{eqnarray}
\mathcal{M}^{\mu\nu} &=& -g \sqrt{\frac{M_\Phi}{N_c}}
\left\{ {\bf k} \cdot \frac{\partial
\psi({\bf p})}{\partial {\bf p}} \delta^{\nu 0} + k_0 \frac{\partial
\psi({\bf p})}{\partial p^j} \delta^{\nu j} \right\} \delta^{\mu i}
\nonumber \\ && \mbox{} \times
\bar{u}(p_1) \frac{1+\gamma_0}{2} \gamma^i \frac{1-\gamma_0}{2}T^a v(p_2).
\label{eq:amp}
\end{eqnarray}

The scattering cross section $\sigma_{\Phi g}$ can be obtained from 
\begin{widetext}
\begin{eqnarray}
\sigma_{\Phi g} &=& \int \frac{1}{4 M_\Phi k_0}
\overline{|\mathcal{M}|^2}
(2\pi)^4 \delta^4(p_1+p_2-k-q) \frac{d^3 p_1}{2p_1^0 (2\pi)^3}
 \frac{d^3 p_2}{2p_2^0 (2\pi)^3}
\nonumber \\
&=& \int \frac{|{\bf p}_1|^3}{ M_\Phi (k_0 p_1^0 - m^2) + m^2(p_1^0 -
k_0) + p_1^0(M_\Phi^2 - 2m^2)/2} \ \frac{\overline{|\mathcal{M}|^2}}{64
\pi^2 M_\Phi k_0} d\, \Omega_{\rm lab}
\nonumber \\
&\approx& \int \frac{\sqrt{k_0+\varepsilon}}{128\pi^2 M_\Phi k_0 \sqrt{m}}
\overline{|\mathcal{M}|^2} d\, \Omega_{\rm lab}
\label{phase}
\end{eqnarray}
\end{widetext}
in the $\Phi$ rest frame, where
$
\overline{|\mathcal{M}|^2} = \frac{1}{6(N_c^2-1)} \sum
|\mathcal{M}|^2
$
and the average is over the initial polarizations and the color of the
gluon.
Note that going from the second line to the third line in
Eq.~(\ref{phase}), we use the non-relativistic approximation
(\ref{counting}).   
With the amplitude (\ref{eq:amp}) we obtain
\begin{equation}
\overline{|\mathcal{M}|^2} = \frac{4g^2 m^2 M_\Phi k_0^2}{3 N_c} \left|
\bm{\nabla} \psi ({\bf p}) \right|^2.
\label{eq:Mfin2}
\end{equation}
Making use of the ground state wave function of the Coulombic bound
state,
\begin{equation}
\bm{\nabla} \psi_{1S}({\bf k}) = -i a_0^{5/2} 32 \sqrt{\pi}
\frac{a_0\bf{k}} {[ (a_0 k)^2 + 1]^3},
\end{equation}
where $\varepsilon = -\varepsilon_0$ and $a_0=16\pi/(g^2N_cm)$,
we finally obtain
\begin{equation}
\sigma_{\Phi g} (\lambda) = \frac{128 g^2}{3 N_c} a_0^2
\frac{(\lambda/\varepsilon_0 - 1)^{3/2}}{(\lambda/\varepsilon_0)^5}
\label{eq:sigma-phig}
\end{equation}
with $\lambda = q \cdot k / M_\Phi$.
Hence, we have derived anew the result for $\sigma_{\Phi g}$ of
Ref. \cite{BP79}.
The scattering cross section $\sigma_{\Phi h}$ can then be computed
using the factorization formula (\ref{eq:fact}).

To apply the same procedure to the scattering of $\Phi'$ (the
$2S$ state), we use the wave function
\begin{equation}
\bm{\nabla} \psi_{2S} ({\bf k}) = -i a_0^{5/2} 8 \sqrt{2\pi}
\frac{[(a_0 k)^2 - \textstyle\frac12]}{[(a_0 k)^2 +
\textstyle\frac14]^4} a_0 {\bf k},
\end{equation}
with the binding energy $\varepsilon = - \varepsilon_0/4$.
This then leads to
\begin{equation}
\sigma_{\Phi' g} = \frac{16 g^2}{N_c^2} a_0^2
\frac{(\lambda/\varepsilon_0 - \textstyle\frac14)^{3/2}
(\lambda/\varepsilon_0 - \textstyle\frac34)^2}
{(\lambda/\varepsilon_0)^7},
\end{equation}
which is consistent with the result of Ref. \cite{AGGA01}.
It should be noted, however, that the application of our formalism should 
be less reliable for the excited states becasue the wave function would 
be more affected by the confining part of the potential.

\begin{figure}[t]
\centerline{\epsfig{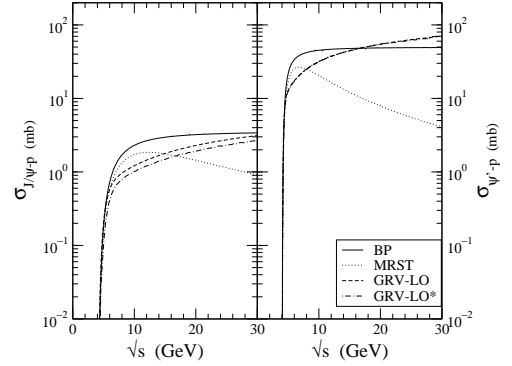}}
\caption{Scattering cross sections $\sigma_{J/\psi-p}$ and
$\sigma_{\psi'(2S)-p}$.
The solid lines are obtained with $g(x)$ of Ref. \cite{BP79} (BP),
while the dashed and dotted lines are with that of
Ref. \cite{MRST98} (MRST) and \cite{GRV95} (GRV-LO), respectively.
The cross sections with relativistic corrections are given by 
dot-dashed lines with $g(x)$ of Ref. \cite{GRV95} (GRV-LO*).}
\label{fig:res1}
\end{figure}

We now plot the numerical results for the $\sigma_{\Phi p}$ and
$\sigma_{\Phi' p}$ in Fig.~\ref{fig:res1}.
Similar plot has been already given in Refs. \cite{KS94,AGGA01}, but here we
use the original parameters as given in Ref. \cite{BP79}:
$\varepsilon_0 = 780 \mbox{ MeV}$ and $m = 1.95 \mbox{ GeV}$
as determined in Ref. \cite{BP79} by fitting the $J/\psi$ and
$\psi'(2S)$ masses to a Coulombic spectrum.
Also given in Fig.~\ref{fig:res1} are the results with the relativistic
correction (dot-dashed lines) that will be discussed below.
For the gluon distribution function of the proton, we use the
parameterization of Ref. \cite{BP79}, $g(x) =  0.5 (\eta+1) (1-x)^\eta/x$
with $\eta =5$ (BP) as well as the MRST \cite{MRST98} and the GRV-LO gluon
distributions \cite{GRV95}.
The results show the dependence of the cross sections on the gluon
distribution function.
With the distribution functions of Refs. \cite{BP79,GRV95}, the cross
sections increase with the center-of-mass energy $\sqrt{s}$.
But $g(x)$ of Ref. \cite{MRST98} leads to the decrease of the cross
sections at higher energies, which
comes from the difference in $g(x)$ at small $x$.
Looking at the  ratio $\sigma_{\psi'(2s)-p}^{} /
\sigma_{J/\psi-p}^{}$ within this approach, one finds that it has a peak at
$\sqrt{s} \approx 4.2$ GeV.
The peak value is about $2000 \sim 5000$ but it quickly converges to
$4 \sim 20$ at higher energies depending on the form of $g(x)$. 
In a quark exchange model the ratio between the maxima of the cross
sections was estimated to be around $6$ \cite{WSB01}.

\begin{figure}[h]
\centerline{\epsfig{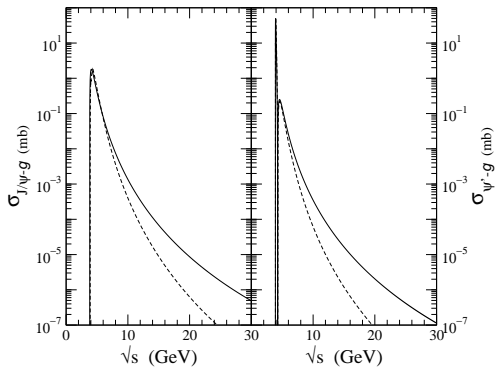}}
\caption{Scattering cross sections $\sigma_{J/\psi-g}$ and
$\sigma_{\psi'(2S)-g}$.
The solid lines are obtained in the non-relativistic limit and the
dashed lines are with the relativistic correction.}
\label{fig:elem}
\end{figure}

Let us now consider the relativistic corrections to our results.   
The non-relativistic counting scheme (\ref{counting}) of this formalism
implies that the formula (\ref{eq:sigma-phig}) should not be used at 
high gluon energy.
However, calculating relativistic corrections in this formalism is
intricately related to the higher $g^2$ corrections, which is a formidable
future task.
Here, we will only look at the simple but important relativistic correction
coming from the relativistic calculation of the phase space integral in
Eq.~(\ref{phase}).
The difference after a full calculation is shown in Fig.~\ref{fig:elem}
for $\sigma_{\Phi g}$, which shows larger suppression of the cross
sections at higher energy.
In order to see the relativistic corrections in $\sigma_{\Phi p}$
we compute $\sigma_{\Phi p}$ with the
$\sigma_{\Phi g}$ after the relativistic correction using the GRV-LO
gluon distribution function (the dot-dashed lines in
Fig.~\ref{fig:res1}).
We find that this relativistic correction does not have much effect
on the cross section $\sigma_{\Phi h}$.  This follows because 
$\sigma_{\Phi g}$ is highly peaked at small energy region and the 
gluon distribution function increases towards small $x$.  
Through Eq.~(\ref{eq:fact}), this implies that 
$\sigma_{\Phi h}$, at all energies, are dominated by the low energy 
behavior of $\sigma_{\Phi g}$, which is reliably calculated in the 
non-relativistic limit.  
In fact, it has been observed in a potential model that the relativistic
treatment of the bound state wave functions does not improve much the
nonrelativistic description \cite{LRS92}.
All these imply that the full relativistic corrections to the scattering
amplitude $\sigma_{\Phi g}$ would have little effects on $\sigma_{\Phi h}$.

The remaining important and interesting correction is the higher twist-effect,
which contributes as $A/\varepsilon^2$ in our formalism. 
Since $A$ is related to some hadronic scale of the hadron $h$, it should
be of the order $\varepsilon^2$ itself, hence should be non-negligible.  
The research in this direction awaits more experimental and theoretical
works \cite{BB00,KL01}.  
It is also important to extend this calculation to investigate the
absortion cross section for the $\chi$ states.
This will influence the amount of $J/\psi$ production coming from the decay
of the $\chi$'s in a p-A or A-A reaction \cite{Ge98}.

% Acknowledgements

We are grateful to A. Hayashigaki and C.-Y. Wong for fruitful discussions.
This work was supported in part by the Brain Korea 21 project of Korean
Ministry of Education, KOSEF under Grant No. 1999-2-111-005-5, the
Yonsei University Research Grant, and the Korean Ministry of Education
under Grant No. 2000-2-0689.


\begin{thebibliography}{10}

\bibitem{Peskin79}
M. E. Peskin,
   Nucl. Phys. {\bf B156}, 365 (1979).
%%CITATION = NUPHA,B156,365;%%

\bibitem{BP79}
G.~Bhanot and M. E. Peskin,
   Nucl. Phys. {\bf B156}, 391 (1979).
%%CITATION = NUPHA,B156,391;%%

\bibitem{ADM78}
T.~Appelquist, M.~Dine, and I.~Muzinich,
   Phys. Rev. D {\bf 17}, 2074 (1978).
%%CITATION = PHRVA,D17,2074;%%

\bibitem{KS94}
D.~Kharzeev and H.~Satz,
   Phys. Lett. B {\bf 334}, 155 (1994);
%D.~Kharzeev, H.~Satz, A.~Syamtomov, and G.~Zinovjev,
D.~Kharzeev {\it et al.\/},
   {\it ibid.\/} {\bf 389}, 595 (1996).
%%CITATION = PHLTA,B334,155;%%
%%CITATION = PHLTA,B389,595;%%

\bibitem{AGGA01}
%F.~Arleo, P.-B. Gossiaux, T.~Gousset, and J.~Aichelin,
F.~Arleo {\it et al.\/},
   Phys. Rev. D {\bf 65}, 014005 (2002).
%%CITATION = HEP-PH 0102095;%%

\bibitem{MS86}
T.~Matsui and H.~Satz,
   Phys. Lett. B {\bf 178}, 416 (1986).
%%CITATION = PHLTA,B178,416;%%

\bibitem{Vogt99}
R.~Vogt,
   Phys. Rep. {\bf 310}, 197 (1999).
%%CITATION = PRPLC,310,197;%%

\bibitem{meson-ex}
S. G. Matinyan and B.~M{\"u}ller,
   Phys. Rev. C {\bf 58}, 2994 (1998);
%%CITATION = PHRVA,C58,2994;%%
K. L. Haglin,
   {\it ibid.\/} {\bf 61}, 031902 (2000);
%%CITATION = PHRVA,C61,031902;%%
Z.~Lin and C. M. Ko,
   {\it ibid.\/} {\bf 62}, 034903 (2000);
%%CITATION = PHRVA,C62,034903;%%
Y.~Oh, T.~Song, and Su H. Lee,
   {\it ibid.\/} {\bf 63}, 034901 (2001);
%%CITATION = PHRVA,C63,034901;%%
A.~Sibirtsev, K. Tsushima, and A. W. Thomas,
   {\it ibid.\/} {\bf 63}, 044906 (2001);
%%CITATION = PHRVA,C63,044906;%%
W.~Liu, C. M. Ko, and Z. W. Lin,
   nucl-th/0107058.
%   Phys.  Rev. C (in press).
%%CITATION = NUCL-TH 0107058;%%

\bibitem{WSB01}
C.-Y. Wong, E. S. Swanson, and T. E. Barnes,
   Phys. Rev. C {\bf 62}, 045201 (2000);
%%CITATION = PHRVA,C62,045201;%%
   {\bf 65}, 014903 (2002);
%%CITATION = NUCL-TH 0106067;%%
C.-Y. Wong,
   {\it ibid.\/} {\bf 65}, 034902 (2002);
   private communications.
%%CITATION = NUCL-TH 0110004;%%

\bibitem{NNMK02}
F.~S. Navarra {\it et al.\/},
  Phys. Lett. B {\bf 529}, 87 (2002)
%%CITATION = PHLTA,B529,87;%%

\bibitem{HVN81}
B.~Humpert and W. L. van Neerven,
   Phys. Lett. B {\bf 102}, 426 (1981);
   Phys. Rev. D {\bf 25}, 2593 (1982).
%%CITATION = PHLTA,B102,426;%%
%%CITATION = PHRVA,D25,2593;%%

\bibitem{HLZ88}
T. H. Hansson, Su~H. Lee and I.~Zahed,
  Phys. Rev. {\bf D37}, 2672 (1988).
%%CITATION = PHRVA,D37,2672;%%

\bibitem{BB00}
%J.~Bartels, C.~Bontus, and H.~Spiesberger,
J.~Bartels {\it et al.\/},
   hep-ph/9908411 (unpublished);
%%CITATION = HEP-PH 9908411;%%
J.~Bartels and C.~Bontus,
   Phys. Rev. D {\bf 61}, 034009 (2000).
%%CITATION = PHRVA,D61,034009;%%

\bibitem{KL01}
S.~Kim and Su~H. Lee,
   Nucl. Phys. {\bf A679}, 517 (2001).
%%CITATION = NUPHA,A679,517;%%

\bibitem{EFP82}
R.~K. Ellis, W. Furmanski and R. Petronzio,  
   Nucl. Phys. {\bf B207}, 1 (1982);
  {\bf B212}, 29 (1983).
%%CITATION = NUPHA,B207,1;%%
%%CITATION = NUPHA,B212,29;%%

\bibitem{BS51}
E. E. Salpeter and H. A. Bethe,
   Phys. Rev. {\bf 84}, 1232 (1951).
%%CITATION = PHRVA,84,1232;%%

\bibitem{EFG00}
%D.~Ebert, R.N. Faustov, and V.O. Galkin,
D.~Ebert {\it et al.\/},
   Phys. Rev. D {\bf 62}, 034014 (2000).
%%CITATION = PHRVA,D62,034014;%%

\bibitem{MRST98}
%A.D. Martin, R.G. Roberts, W.J. Stirling, and R.S. Thorne,
A. D. Martin {\it et al.\/},
   Eur. Phys. J. C {\bf 4}, 463 (1998).
%%CITATION = EPHJA,C4,463;%%

\bibitem{GRV95}
%M.~Gl{\"u}ck, E.~Reya, and A.~Vogt,
M.~Gl{\"u}ck {\it et al.\/},
   Z. Phys. C {\bf 67}, 433 (1995).
%%CITATION = ZEPYA,C67,433;%%

\bibitem{LRS92}
%W.~Lucha, H.~Rupprecht, and F.F. Sch{\"o}berl,
W.~Lucha {\it et al.\/},
   Phys. Rev. D {\bf 46}, 1088 (1992).
%%CITATION = PHRVA,D46,1088;%%

\bibitem{Ge98}
L. Gerland {\it et al.\/},
  Phys. Rev. Lett. {\bf 81}, 762 (1998)
%%CITATION = PRLTA,81,762;%%

\end{thebibliography}
\end{document}